\documentclass[final]{aipproc}

\layoutstyle{6x9}

\begin{document}

\title{Next-to-leading Order QCD Corrections to $A_{\mathrm{TT}}$ for 
Single-Inclusive Hadron Production \\[-2.6cm]
\hspace*{10.2cm}{\small{BNL-NT-05/23, 
RBRC-531 \\}}
\vspace*{2.5cm}}


\author{A. Mukherjee}{
  address={Instituut-Lorentz, University of Leiden, 2300 RA Leiden, 
The Netherlands}}

\author{M. Stratmann}{
  address={Institut f{\"u}r Theoretische Physik, Universit{\"a}t
Regensburg, D-93040 Regensburg, Germany}}

\author{W. Vogelsang}{
  address={Physics Department and RIKEN-BNL Research Center,    
Brookhaven National Laboratory, Upton, New York 11973, U.S.A.} }

\begin{abstract}
We report on  a calculation of the next-to-leading order QCD corrections
to the partonic cross sections contributing to single-inclusive high-$p_T$
hadron production in collisions of transversely polarized hadrons. We give 
some predictions for the double spin asymmetry $A_{\mathrm{TT}}^{\pi}$ for 
the proposed experiments at RHIC and at the GSI.
\end{abstract}

\maketitle

\section{Introduction}

The leading-twist partonic structure of a spin-${1\over 2}$    
hadron is given in terms of the unpolarized parton distribution functions 
$f(x,Q^2)$, the helicity distributions $\Delta f (x, Q^2)$, 
and the transversity distributions $\delta f(x, Q^2)$. Transversity
describes the number density of a parton with the same transverse polarization 
as the nucleon, minus the number density for opposite polarization. Among
the various parton distributions, the $\delta f(x,Q^2)$ are the ones
about which we have the least knowledge. They are at present the focus of 
much experimental activity. Transversity will be probed by double-transverse 
spin asymmetries in transversely polarized $pp$ collisions at the BNL 
Relativistic Heavy Ion  Collider (RHIC) \cite{rhic}. 
The most promising reaction is the Drell-Yan process, 
which offers the largest spin asymmetries, but
whose main drawback is the rather moderate event rate \cite{martin}. 
Other relevant processes include high $p_T$ prompt photon \cite{attlo,photonnlo}
and jet production~\cite{attlo}. However, for these the
asymmetry is expected to be much smaller because of the absence of
gluon-initiated subprocesses in the transversely polarized cross section. 
Recently, it has also been investigated whether one could extract transversity 
from measurements of $A_{\mathrm{TT}}$ for the Drell-Yan 
process in transversely polarized 
${\bar p}p$ collisions at the planned GSI-FAIR 
facility~\cite{pax,assia,anselmino,efremov,resum,radici}.
In this note (for details, see Ref.~\cite{pion}), we consider the spin asymmetry
in single-inclusive production of pions
at large transverse momentum $p_T$ as a possible means for 
determining transversity at RHIC or the GSI. We report on the calculation
of the next-to-leading order (NLO) corrections for this reaction, and also present
some phenomenological results.

\section{Projection Technique}

It is known that the NLO QCD corrections are required
in order to have firm theoretical predictions for hadronic scattering. Only with 
their knowledge can one reliably extract information on the 
partonic (spin) structure of nucleons.
Apart from this motivation, also interesting new technical questions
arise beyond leading order (LO) in the calculations of cross sections 
with transverse 
polarization. Unlike the longitudinally polarized case, where the spin
vectors are aligned with the momenta, the transverse spin vectors specify
extra spatial directions and, as a result, the cross section has non-trivial
dependence  on the azimuthal angle of the observed particle. For $A_{TT}$  
this dependence is of the form \cite{photonnlo} 
\begin{eqnarray}
\label{eq2}
\frac{d^3\delta \sigma}{dp_T d\eta d\Phi}\;\equiv\;
\cos (2\Phi)\,\left\langle \frac{d^2\delta\sigma}{dp_T d\eta}\right\rangle
\; ,
\end{eqnarray}
for a parity conserving theory with vector coupling. Here, the spin vectors 
are taken to point in the $\pm x$ direction. We furthermore consider the 
scattering in the center-of-mass frame of the initial hadrons and use their
momenta to define the $z$ axis. Because of the $\cos(2\Phi)$ dependence, 
integration over all azimuthal angles is not appropriate. This
makes it difficult to use the standard techniques developed for NLO
calculations of unpolarized and longitudinally polarized processes, because these
techniques usually rely on the integration over the full azimuthal phase 
space and also on the choice of particular reference frames that are related in
complicated ways to the center-of-mass frame of the initial hadrons. In
\cite{photonnlo} a general technique was introduced that facilitates NLO
calculations for transverse polarization by conveniently projecting on the
azimuthal dependence of the cross section in a covariant way. The projector 
\begin{eqnarray}
\label{eq5}
F(p,s_a,s_b) =
\frac{s}{\pi t u}
\,\left[ 2 \,(p\cdot s_a)\, (p\cdot s_b)\; +\;
\frac{t u}{s} \,(s_a \cdot s_b) \right]
\end{eqnarray}
reduces to $\cos (2 \Phi)/\pi$ in the center-of-mass frame of the
initial hadrons. Here $p$ is the momentum of the observed particle in the
final state and the $s_i$ are the initial transverse spin vectors. 
The squared matrix element for the partonic process is multiplied with this 
projector and integrated over the full azimuthal phase space. Integrations of 
terms involving the product of the transverse spin vectors with the 
final-state momenta can be performed using a tensor decomposition. After this step, 
no scalar products involving the $s_i$ are left in the squared matrix element. For
the remainder of the phase space integrations, one can now use techniques familiar 
from the unpolarized and longitudinally polarized cases. This method is
particularly convenient at NLO, where one uses dimensional regularization and
the phase space integrations are performed in $n\neq 4$ dimensions.

\section{Application to single-inclusive hadroproduction : 
phenomenological results}

\begin{figure}
  \includegraphics[width=5cm]{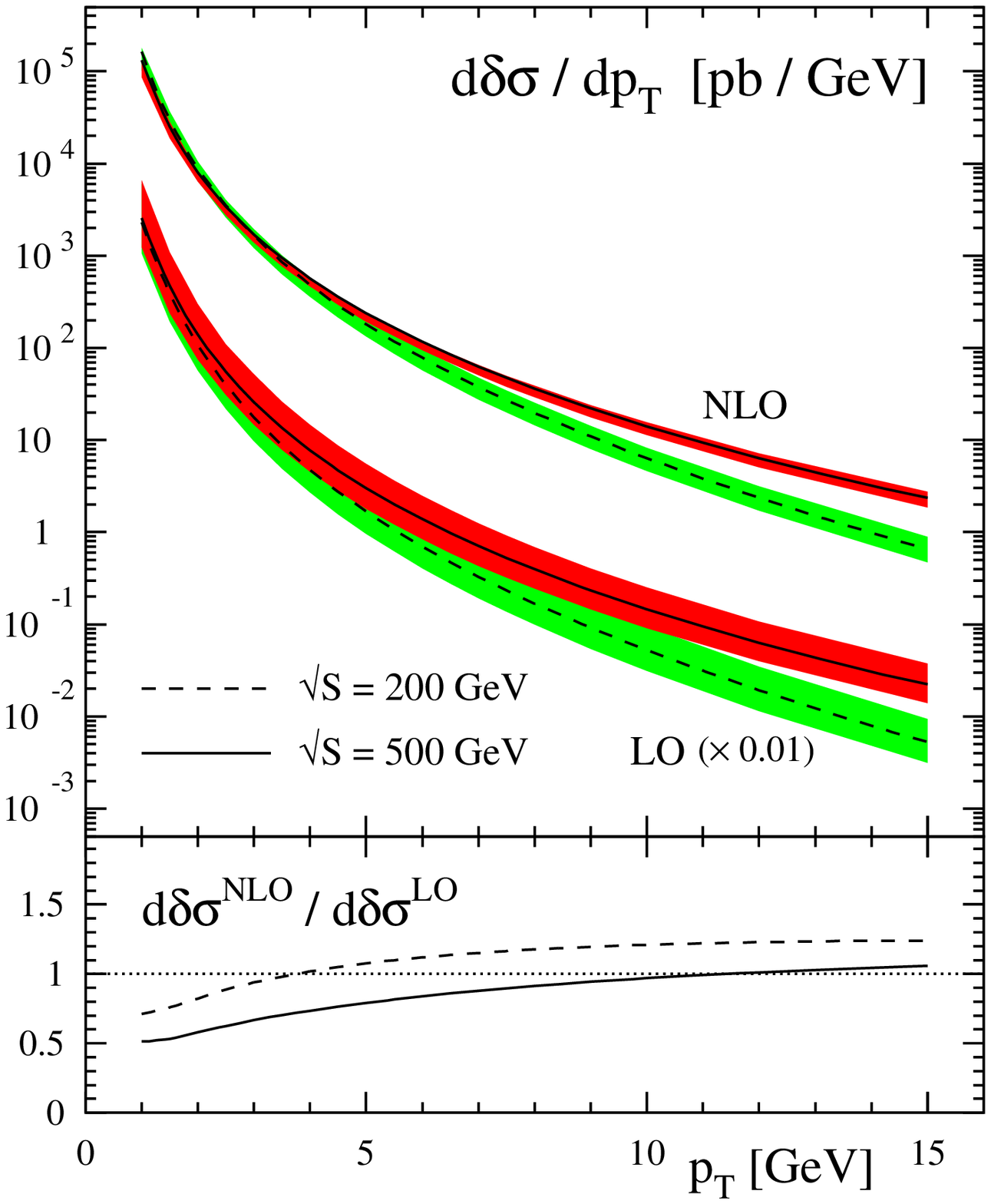}
\hspace*{6mm}
\includegraphics[width=6cm]{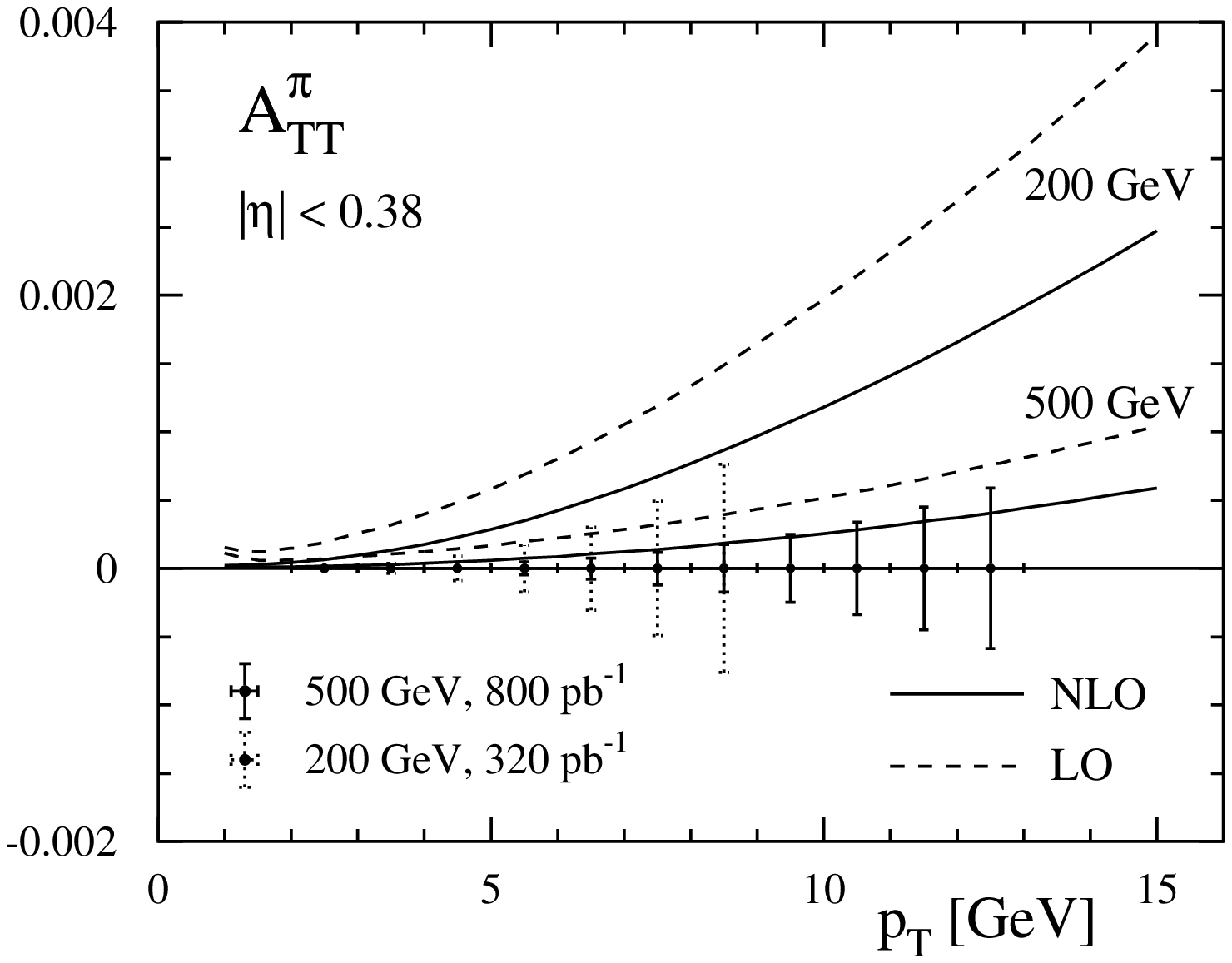}

\caption{\label{fig1} 
Predictions for the transversely polarized single-inclusive pion
production cross sections at LO and NLO at RHIC (left), and for the transverse 
double-spin asymmetry $A_{\mathrm{TT}}^{\pi^0}$ (right). 
The shaded bands represent the range of predictions when the 
scale $\mu$ is varied in the range $p_T \le \mu \le 4p_T$. 
The lower panel on the left shows the ratios of the NLO and LO results (the
``$K$-factors'').}
\end{figure}

Next, we give some phenomenological results for transversely polarized  
$pp$ collisions at RHIC ($\sqrt{S}=200$ and 500~GeV) and $\bar{p}p$
collisions at the GSI-FAIR facility in an asymmetric collider mode 
with proton and 
antiproton energies of $3.5$ GeV and $15$~GeV, respectively \cite{pion}.
For our numerical predictions, we model the transversity distributions by
saturating the Soffer inequality~\cite{soffer} at some low input scale
$\mu_0 \approx 0.6 $ GeV, using 
the NLO (LO) GRV \cite{grv} and GRSV (``standard scenario'')
\cite{grsv} densities $q(x,\mu_0)$ and $\Delta q(x,\mu_0)$, 
respectively.

Figure~\ref{fig1} (left) shows our predictions
for the transversely polarized single-inclusive pion production cross sections
at  LO and NLO for the two different c.m.s.\ energies at RHIC. We also 
display the scale uncertainty. For the $K$-factor shown in the lower panel the scale 
choice is $\mu=2 p_T$. A significant decrease
of scale dependence is observed when going from LO to NLO.
 Figure~\ref{fig1} (right) shows the asymmetry
$A_{\mathrm{TT}}^{\pi^0}$ defined as the ratio of the polarized and unpolarized 
cross sections. Here, the scale is set to $\mu=p_T$. We also 
display the statistical errors
that may be achievable in experiment. We have calculated these using
\begin{equation} \label{error}
\delta A_{\mathrm{TT}}
\simeq \frac{1}{P_1 P_2 \sqrt{{\cal L}\,\sigma_{\rm bin}}} \; ,
\end{equation}
where $P_1=P_2=0.7$  are the transverse 
polarizations of the proton beams, ${\cal L}$ the integrated
luminosity of the collisions indicated in the figure, 
and $\sigma_{\rm bin}$ the unpolarized
cross section integrated over the $p_T$-bin for which the error is to be
determined. The asymmetry is very small.

\begin{figure}
  \includegraphics[width=5cm]{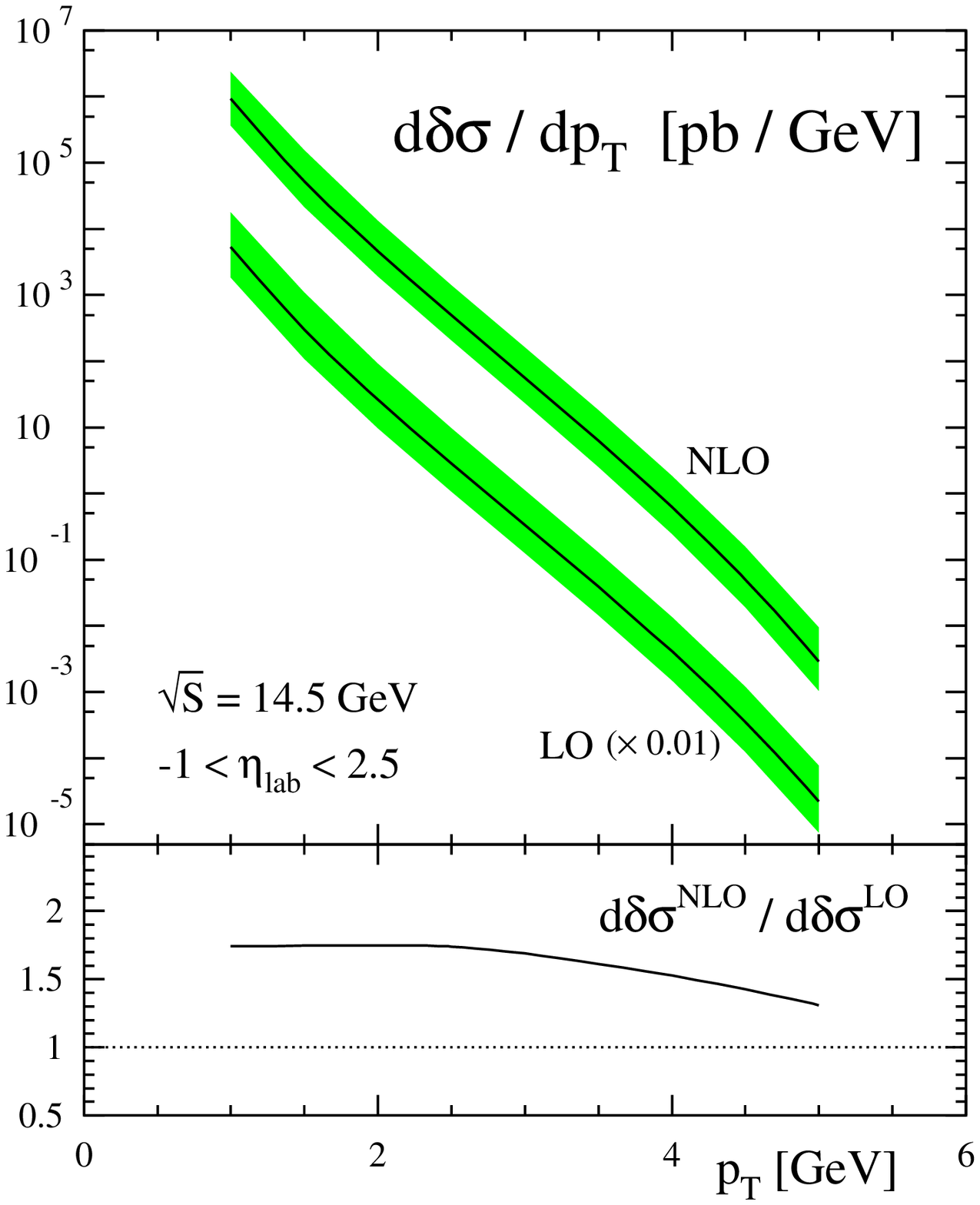}
\hspace*{6mm}
\includegraphics[width=6cm]{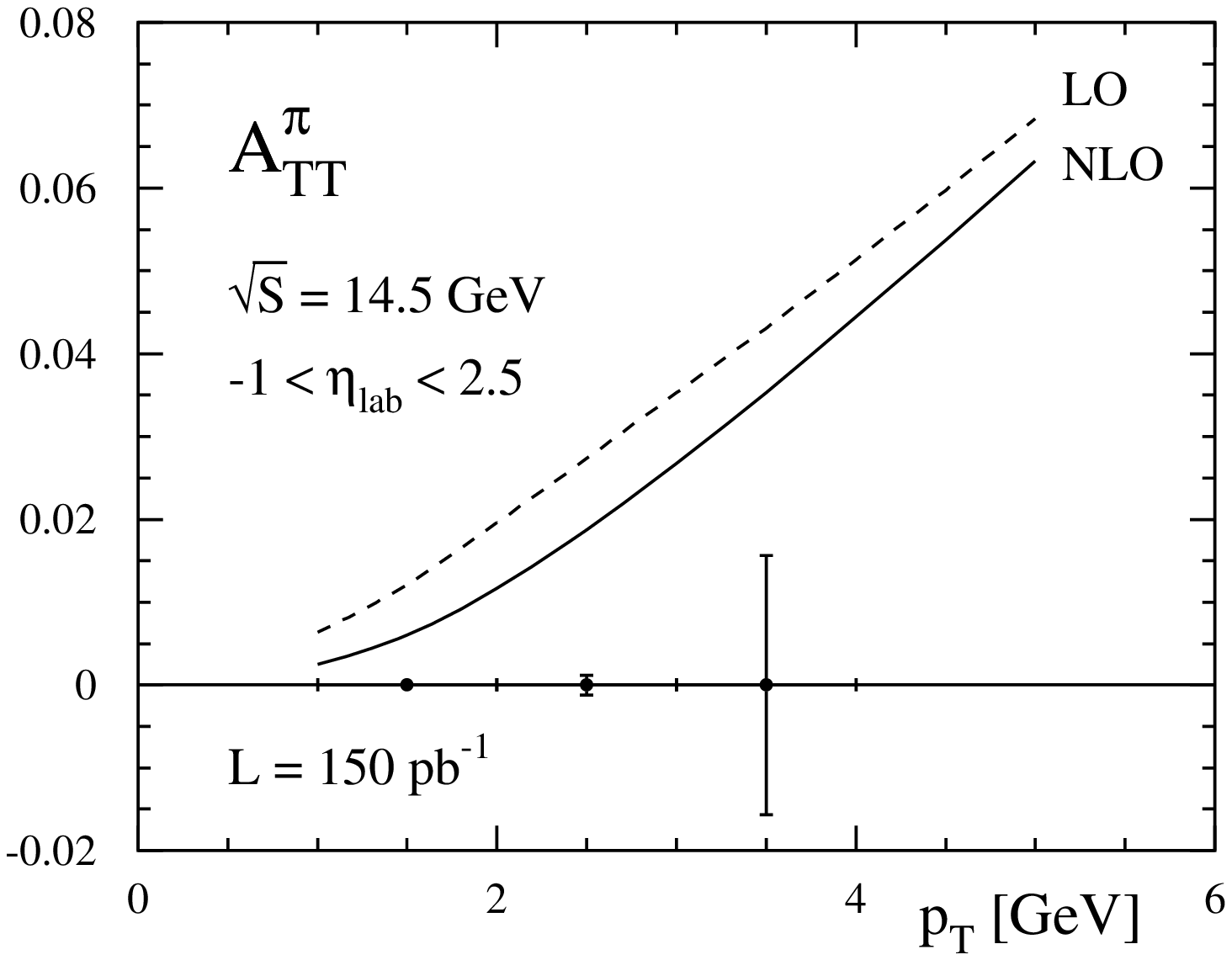}
\caption{\label{fig2} LO and NLO predictions for the cross section (left) and the
transverse spin asymmetry $A_{\mathrm{TT}}^{\pi^0}$ (right)
for single-inclusive pion production in $\bar{p}p$ collisions at the GSI. 
The shaded bands represent the range of predictions if the 
scale $\mu$ is varied in the range $p_T \le \mu \le 4p_T$. 
The lower panel (left) shows the ratios of the NLO and LO results (the
``$K$-factors'').} 
\end{figure}

Figure~\ref{fig2} (left) shows the corresponding  
cross sections in transversely polarized $\bar{p}p$ collisions  
at $\sqrt{S}= 14.5$~GeV in an asymmetric collider mode at the GSI-FAIR facility.
We have  assumed beam polarizations of $30\%$ and $50 \%$ for the 
antiprotons and  protons, respectively.   
At GSI energies, the scale dependence does not really improve from LO to NLO.  
A resummation of the double-logarithmic corrections to the partonic cross
sections~\cite{ref:resum} 
would be very desirable for the future, along with a study of
power corrections. Figure~\ref{fig2} (right) shows the predicted transverse
double-spin asymmetry at the GSI, which is much larger than at RHIC.  

\begin{theacknowledgments}
A.M. thanks the organizers of DIS 2005 for a wonderful conference and the
invitation.  W.V.\ is grateful to RIKEN, Brookhaven National Laboratory and
the Department of Energy (contract number DE-AC02-98CH10886) for
providing the facilities essential for the completion of this work.
This work is supported in part by the ``Bundesministerium f\"{u}r Bildung und
Forschung (BMBF)'' and by FOM, The Netherlands.
\end{theacknowledgments}



\begin{thebibliography}{99}
%
\bibitem{rhic} See, for example: G.\ Bunce, N.\ Saito, J.\ Soffer, and
W.\ Vogelsang, Annu.\ Rev.\ Nucl.\ Part.\ Sci.\ {\bf 50}, 525 (2000).
%
\bibitem{martin} O.\ Martin, A.\ Sch\"{a}fer, M.\ Stratmann, and 
W.\ Vogelsang,  Phys. Rev. {\bf D57}, 3090 (1998); {\bf D60}, 117502 (1999).
%
\bibitem{attlo} J.\ Soffer, M.\ Stratmann, and W.\ Vogelsang,
Phys. Rev. {\bf D65}, 114024 (2002).
%
\bibitem{photonnlo} A. Mukherjee, M. Stratmann, and W. Vogelsang, Phys.
Rev. {\bf D67}, 114006 (2003).
%
\bibitem{pax} P.~Lenisa and F.~Rathmann  [the PAX Collaboration],
{\tt hep-ex/0505054} and {\tt http://www.fz-juelich.de/ikp/pax/}
%
\bibitem{assia} M.~Maggiora  [the ASSIA Collaboration], {\tt hep-ex/0504011};
GSI-ASSIA Technical Proposal, Spokesperson: R. Bertini,
{\tt http://www.gsi.de/documents/DOC-2004-Jan-152-1.ps}
%
\bibitem{anselmino}
M.~Anselmino, V.~Barone, A.~Drago, and N.~N.~Nikolaev,
Phys.\ Lett.\ {\bf B594}, 97 (2004).
%
\bibitem{efremov} A.~V.~Efremov, K.~Goeke, and P.~Schweitzer,
Eur.\ Phys.\ J.\ {\bf C35}, 207 (2004). 
%
\bibitem{resum} H. Shimizu, G. Sterman, W. Vogelsang, and H. Yokoya,
Phys.\ Rev.\ {\bf D71}, 114007 (2005).
%
\bibitem{radici} A.~Bianconi and M.~Radici, {\tt hep-ph/0504261}.
%
\bibitem{pion}  A. \ Mukherjee, M.\ Stratmann, and
W.\ Vogelsang, {\tt hep-ph/0506315}.
%
\bibitem{soffer}  J.\ Soffer, Phys. Rev. Lett. {\bf 74}, 1292 (1995);
D.\ Sivers, Phys. Rev. {\bf D51}, 4880 (1995).
%
\bibitem{grv} M.\ Gl\"{u}ck, E.\ Reya, and A.\ Vogt,
Eur.\ Phys.\ J.\ {\bf C5}, 461 (1998).
%
\bibitem{grsv}  M.\ Gl\"{u}ck, E.\ Reya, M.\ Stratmann, and
W.\ Vogelsang, Phys. Rev. {\bf D63}, 094005 (2001).
%
\bibitem{ref:resum} D.~de Florian and W.~Vogelsang, Phys. Rev. {\bf D71},
114004 (2005).
%
\end{thebibliography}
\end{document}